\documentclass[conference, 9pt, a4paper]{IEEEtran}
\IEEEoverridecommandlockouts

\usepackage{cite}
\usepackage{amsmath,amssymb,amsfonts}
\usepackage{algorithmic}
\usepackage{graphicx}
\usepackage{textcomp}
\usepackage{xcolor}
\usepackage{pgfplots}
\usepackage{float}
\usepackage{balance}
\pgfplotsset{compat=1.18}
\usepackage{tikz}
\usetikzlibrary{arrows.meta, positioning, calc, patterns}

\def\BibTeX{{\rm B\kern-.05em{\sc i\kern-.025em b}\kern-.08em
    T\kern-.1667em\lower.7ex\hbox{E}\kern-.125emX}}
\begin{document}
\title{TroPUF: Evaluating Hardware Trojan Insertion in Delay-Based Physical Unclonable Functions\\}

\author{\IEEEauthorblockN{Marissa Marcarelli}
\IEEEauthorblockA{\textit{Computer Engineering \& Computer Science Department} \\
\textit{California State University Long Beach}\\
Long Beach, CA, USA \\
marissa.marcarelli01@student.csulb.edu}
\and
\IEEEauthorblockN{Amin Rezaei}
\IEEEauthorblockA{\textit{Computer Engineering \& Computer Science Department} \\
\textit{California State University Long Beach}\\
Long Beach, CA, USA \\
amin.rezaei@csulb.edu}
}

\maketitle

\begin{abstract}

Delay-based Physical Unclonable Functions (PUFs) are commonly used for device authentication and key generation due to the fact that they rely on manufacturing induced delay variations. However, these same variations make PUFs inherently non-deterministic, which can allow malicious logic to blend in with normal circuit behavior. As a result, the act of embedding hardware Trojans directly inside the PUF primitive presents a unique security risk that is not yet well understood. This work presents a unified simulation framework for evaluating stealthy hardware Trojan insertion across multiple delay-based PUF architectures and Trojan types. Functional metrics, hardware overhead, and resistance to machine learning modeling are assessed in parallel. Results show that dormant Trojans preserve expected PUF behavior, structural characteristics, and modeling resistance. Detectable degradation appears only after activation, indicating that conventional validation techniques fail to identify embedded Trojans prior to payload execution. These findings expose a gap in current PUF security assumptions, and highlight the need to evaluate PUFs and hardware Trojans as a coupled security problem. \\

\end{abstract}


\section{Introduction}

With the increasing reliance on hardware-based authentication and secure key storage, guaranteeing the trustworthiness of security primitives has become progressively important \cite{che2015puf}.

Physical Unclonable Functions (PUFs) \cite{gao2020physical} are hardware security primitives that exploit manufacturing-induced physical variation to generate responses that are difficult to clone and unique between devices \cite{gassend2002silicon,suh2007physical,gao2020physical}. Their quality is commonly evaluated using metrics such as uniqueness, uniformity, randomness, and reliability \cite{maiti2012systematic,hemavathy2023arbiter}, which measure inter-device distinctiveness and intra-device stability. Satisfying these properties is generally considered a foundational requirement for security application use \cite{su2021survey}.

Among different PUFs, delay-based PUFs are one of the most widely studied due to their simplicity and scalability \cite{suh2007physical,wang2019theoretical}. They derive responses by relying on small timing differences along paths designed to be symmetric, where manufacturing-induced variation produces imbalances \cite{ruhrmair2009foundations}.

At the same time, hardware Trojans remain a serious threat in integrated circuits \cite{tehranipoor2010survey,xue2020ten}. A Trojan consists of a rare trigger and malicious payload, and is designed to remain dormant under normal operation \cite{karri2011trojan,jin2009experiences}. Most detection techniques assume deterministic logic and attempt to identify structural, functional, or side-channel anomalies \cite{bhasin2015survey,gubbi2023hardware}. PUFs challenge these assumptions because PUF behavior is intentionally nondeterministic \cite{armknecht2016towards}. Since variability is expected, small irregularities introduced by a Trojan may appear indistinguishable from natural PUF behavior, allowing it to be hidden by the primitive itself.

Machine Learning (ML) further complicates this issue. Modeling attacks have demonstrated that delay-based PUF responses can be approximated with high accuracy using supervised learning techniques \cite{ruhrmair2010modeling,tobisch2015scaling}. While modeling resistance is often treated as evidence of PUF strength \cite{vijayakumar2016machine,ebrahimabadi2021novel}, a Trojan-infected PUF may still preserve modeling resistance under conventional evaluation methods.

\begin{figure*}[t]
\centering
\includegraphics[width=\textwidth]{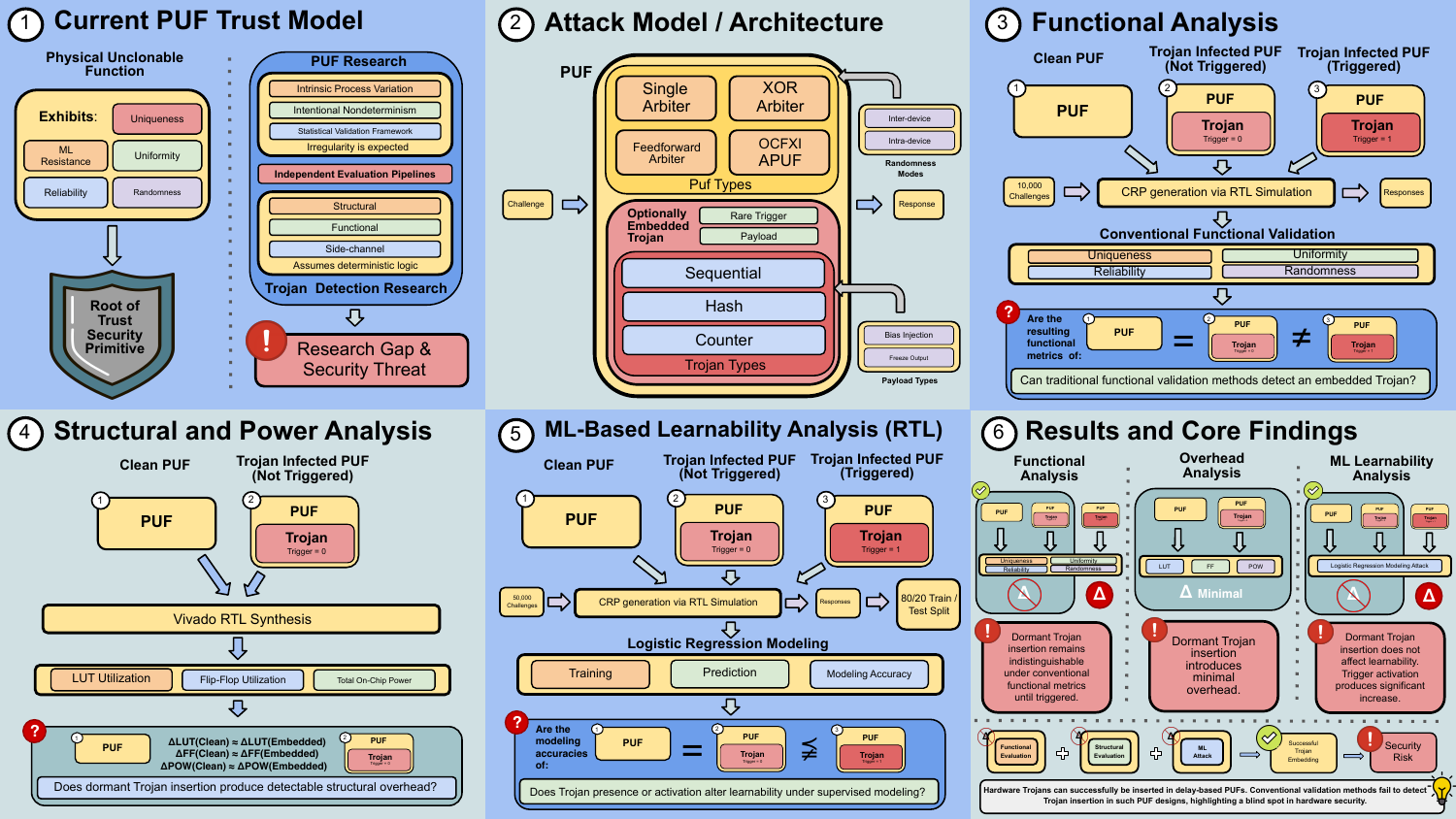}
\caption{TroPUF Framework.}
\label{fig:big_picture}
\end{figure*}

Contrary to common evaluation practices, which treat PUF security and hardware Trojan detection separately, the interaction between the two has not yet been studied in depth. Satisfying functional metrics and modeling resistance does not guarantee integrity when malicious logic is embedded directly within the PUF primitive. In this paper, we present for the first time a unified framework called \textbf{TroPUF} that analyzes hardware Trojan insertion directly in delay based PUFs.
Our contributions, as seen in Fig.~\ref{fig:big_picture} are as follows:
\begin{itemize}
    \item Developing a unified framework for evaluating hardware Trojan insertion across multiple delay-based PUFs.
    \item Analyzing functional metrics and hardware overhead of clean and Trojan-infected PUF implementations.
    \item Evaluating modeling resistance of Trojan-infected PUFs and demonstrating that satisfying conventional PUF metrics and modeling resistance does not guarantee security.
\end{itemize}


\subsection{Related Works} 
\subsubsection{Delay-Based PUFs}
Silicon physical random functions were introduced in earlier work \cite{gassend2002silicon} and later extended for authentication and key generation \cite{suh2007physical,gassend2004identification}. Delay-based PUFs, particularly the Arbiter PUF, exploit race conditions along symmetric delay paths to generate responses from manufacturing variation \cite{suh2007physical,wang2019theoretical}. Architectural extensions include Butterfly \cite{kumar2008butterfly}, Ring Oscillator \cite{sahoo2015novel,chen2011bistable}, XOR \cite{zhou2017secure}, Feed-forward and Interpose \cite{avvaru2019feed,nguyen2018interpose}, and cyclic designs such as CycPUF \cite{dominguez2024cycpuf}. However, increased structural complexity does not inherently improve security \cite{becker2015gap,wisiol2020splitting}. Evaluation typically relies on statistical metrics including uniqueness, reliability, and uniformity \cite{maiti2012systematic,hemavathy2023arbiter,su2021survey}, assuming trusted implementation.

\subsubsection{Modeling Attacks on PUFs}
It was demonstrated that Arbiter PUFs can be approximated using linear additive delay models trained on challenge-response pairs \cite{ruhrmair2010modeling}. Increasing architectural complexity does not inherently prevent learnability \cite{tobisch2015scaling}. This led to the development of ML-resistant designs \cite{vijayakumar2016machine}, modified arbiter structures \cite{ebrahimabadi2021novel}, and robustness evaluation tools such as PUFmeter \cite{ganji2019pufmeter}. Modeling resistance, however, is typically evaluated under benign structural assumptions.

\subsubsection{Hardware Trojans}
Hardware Trojans are typically categorized based on their trigger mechanisms and payload characteristics \cite{tehranipoor2010survey,xue2020ten,karri2011trojan}. Prior work has shown that stealthy insertion is achievable with minimal hardware footprint \cite{jin2009experiences}, including designs that use counter-based triggers \cite{liu2011design}, sequential activation strategies \cite{wang2011sequential}, and scalable mechanisms that exploit rare trigger conditions \cite{lyu2021scalable}. 

\subsubsection{Trojan Detection}
Detection approaches typically rely on structural, functional, and side-channel analysis \cite{bhasin2015survey}. Functional methods attempt rare trigger activation \cite{lyu2021scalable}, while side-channel techniques analyze power and delay deviations \cite{tiwari2023efficient}. ML has also been applied to side-channel data to improve detection accuracy \cite{gubbi2023hardware}. However, these methods assume deterministic circuit behavior. Recent work has further explored advanced detection and strategies, including risk-aware and explainable frameworks that aim to improve coverage and interpretability \cite{vishwakarma2023risk}. Additionally, multimodal deep learning approaches have been proposed to enhance detection performance under uncertainty by leveraging multiple data sources \cite{vishwakarma2024uncertainty}. Complementary efforts have also investigated runtime mitigation techniques for addressing hardware Trojans in deployed systems, particularly in the context of logic-locked circuits \cite{maynard2024reconfigurable}.

\subsubsection{PUFs and Trojan Interaction}
PUFs violate deterministic assumptions due to intrinsic process variation \cite{armknecht2016towards}, complicating anomaly-based detection. Prior work has used PUFs within broader security frameworks \cite{armknecht2016towards} and examined PUF-assisted Trojan attacks \cite{siddik2023puf}. However, these studies treat PUFs primarily as detection or authentication mechanisms. Prior research has not specifically analyzed the security implications of embedding hardware Trojans directly within delay-based PUF primitives, nor evaluated its effects.

\subsection{Motivation}
Delay-based PUF research emphasizes statistical quality and modeling resistance \cite{maiti2012systematic,hemavathy2023arbiter}, while Trojan research focuses on stealth and anomaly detection under deterministic assumptions \cite{tehranipoor2010survey,xue2020ten}. These domains remain largely independent. Whether a Trojan-infected delay-based PUF can simultaneously satisfy functional metrics, preserve modeling resistance, and evade detection remains unexplored. The assumption of trustworthiness does not hold when the security primitive itself serves as the Trojan host. Since delay-based PUFs often serve as root-of-trust primitives, compromise of the PUF also compromises higher level security mechanisms that depend on it. This then enables attacks across entire platforms or systems. This gap motivates and clearly highlights the need for the unified evaluation framework proposed in this work.

\subsection{Threat Model}
We consider an adversary that embeds a hardware Trojan within a delay-based PUF during design or fabrication. The Trojan consists of a rare trigger and dormant payload, remaining inactive under normal operation. Prior to activation, the PUF preserves expected functional and structural behavior and modeling resistance. Upon activation, the payload alters PUF behavior by exploiting intrinsic variability. The defender evaluates the PUF using conventional validation techniques. No trusted reference is assumed, and PUF variability is expected.

\section{Preliminaries}
This section presents the preliminaries relevant to this work, including delay-based PUF architectures, standard evaluation metrics, synthesis-level analysis, and ML-based modeling attacks.

\subsection{Delay Based PUFs}
Delay-based PUFs derive responses from small differences in signal propagation time along nominally symmetric circuit paths. These differences come from manufacturing variation at the transistor and interconnect level, producing device specific imbalances \cite{gassend2002silicon,ruhrmair2009foundations}. Since the delays are physically embedded, the resulting responses are difficult to replicate across devices. The behavior can be modeled as a linear function of the applied challenge, allowing scalable and lightweight implementations. In addition, due to their low area overhead and CMOS compatibility, delay-based PUFs are widely used in authentication and root-of-trust applications\cite{gao2020physical,su2021survey}.

A classic example is the Arbiter PUF, where two signals race through multiplexed delay chains controlled by a challenge vector, and an arbiter outputs a bit based on arrival time \cite{suh2007physical,wang2019theoretical}. Other constructions, such as Ring Oscillator and XOR Arbiter PUFs, exploit frequency differences or combine multiple delay paths to increase complexity \cite{sahoo2015novel,zhou2017secure}.

\subsection{PUF Metrics}
Functional metrics represent the primary means by which PUF designs are traditionally validated and trusted \cite{maiti2012systematic,hemavathy2023arbiter,su2021survey}. By applying the same metrics to both clean and Trojan infected designs, we can determine if standard validation techniques are sufficient to detect malicious modification. In the following formulas, $N$ represents the number of PUF instances, $n$ is the response length in bits, $m$ is the number of evaluated responses, and $s$ is the number of repeated measurements. $R_i$ and $R_j$ are response vectors from different PUF instances, $R_i^*$ represents repeated evaluations of the same PUF instance under identical challenges, and $p(x)$ denotes the probability of observing the response bit $x$.

For reference, the Hamming Distance (HD) between two response vectors $R_1$ and $R_2$ is defined as:
\begin{equation}
\mathrm{HD}(R_1, R_2) = \sum_{i=1}^{n} \left( R_1[i] \oplus R_2[i] \right),
\end{equation}

The Hamming Weight (HW) of a response vector $R$ is given by:
\begin{equation}
\mathrm{HW}(R) = \sum_{i=1}^{n} R[i].
\end{equation}

\subsubsection{Uniqueness}
Different PUF instances with the same design should ideally produce distinct responses to the same challenge. Uniqueness measures this difference, typically defined as the normalized inter-chip HD. The ideal uniqueness that a PUF design can achieve is 50\%.

  \begin{equation}
  \text{Uniqueness} =
  \frac{2}{N(N-1)}
  \sum_{i=1}^{N-1}
  \sum_{j=i+1}^{N}
  \frac{HD(R_i, R_j)}{n}
  \times 100\%
  \end{equation}

\subsubsection{Reliability}
Reliability measures the consistency of a PUF response when the same challenge is applied multiple times. It quantifies the intra-device stability of the generated responses under varying conditions. An ideal PUF achieves a reliability score of 100\%.

\begin{equation}
\text{Reliability} =
\left(
1 -
\frac{1}{s}
\sum_{i=1}^{s}
\frac{HD(R, R_i^*)}{n}
\right)
\times 100\%
\end{equation}

\subsubsection{Uniformity}
Uniformity measures whether the ``1''s and ``0''s in PUF responses are evenly balanced. Ideally, a PUF should produce equal numbers of both values across all challenges, indicating no inherent bias. An ideal PUF achieves a uniformity score of 50\%.

\begin{equation}
\text{Uniformity} =
\frac{1}{m}
\sum_{i=1}^{m}
\frac{HW(R_i)}{n}
\times 100\%,
\end{equation}

\subsubsection{Randomness}
Randomness captures the statistical unpredictability of PUF responses. Ideally, output bits should exhibit no discernible patterns or correlations, even when challenges are closely related. An ideal PUF will reach a randomness score of 100\%.

\begin{equation}
H(X) =
-\sum_{x \in \{0,1\}} p(x)\log_2 p(x) \times 100\%,
\end{equation}

\subsection{Synthesis Analysis}
FPGA synthesis level metrics are evaluated to provide a relative estimate of the structural overhead introduced by Trojan insertion under consistent synthesis conditions \cite{jin2009experiences,tehranipoor2010survey}. These metrics capture changes in resource utilization and power characteristics that may result from embedded malicious logic, including Lookup Table (LUT) utilization, Flip Flop (FF) utilization, and power consumption.

\subsection{ML Modeling Attacks}
ML attacks approximate the functional relationship between challenges and responses using supervised models trained on collected Challenge–Response Pairs (CRPs). For delay-based PUFs such as the Arbiter PUF, responses can be represented using an additive linear delay model, where each challenge corresponds to a feature vector and the output is determined by the sign of a weighted sum \cite{ruhrmair2010modeling}. Algorithms including logistic regression and support vector machines can approximate this model given sufficient CRPs \cite{ruhrmair2010modeling,tobisch2015scaling}. Modeling resistance is commonly evaluated by measuring prediction accuracy under such attacks \cite{vijayakumar2016machine,ganji2019pufmeter}. In this work, prediction accuracy is used as an evaluation metric to assess whether Trojan insertion alters learnability without disrupting functional correctness.

\section{System Architecture and Implementation}

This section describes the proposed unified evaluation framework called \textbf{TroPUF}, which analyzes delay-based PUFs under both clean and Trojan-infected configurations. \textbf{TroPUF} includes various delay-based PUF designs, hardware Trojan models, and payload types to evaluate how malicious logic interacts with PUF behavior across complex implementations.

\subsection{PUF Implementations}

\subsubsection{Single Arbiter PUF (SA-PUF)}
\begin{figure}[H]
\centering
\includegraphics[width=0.9\columnwidth]{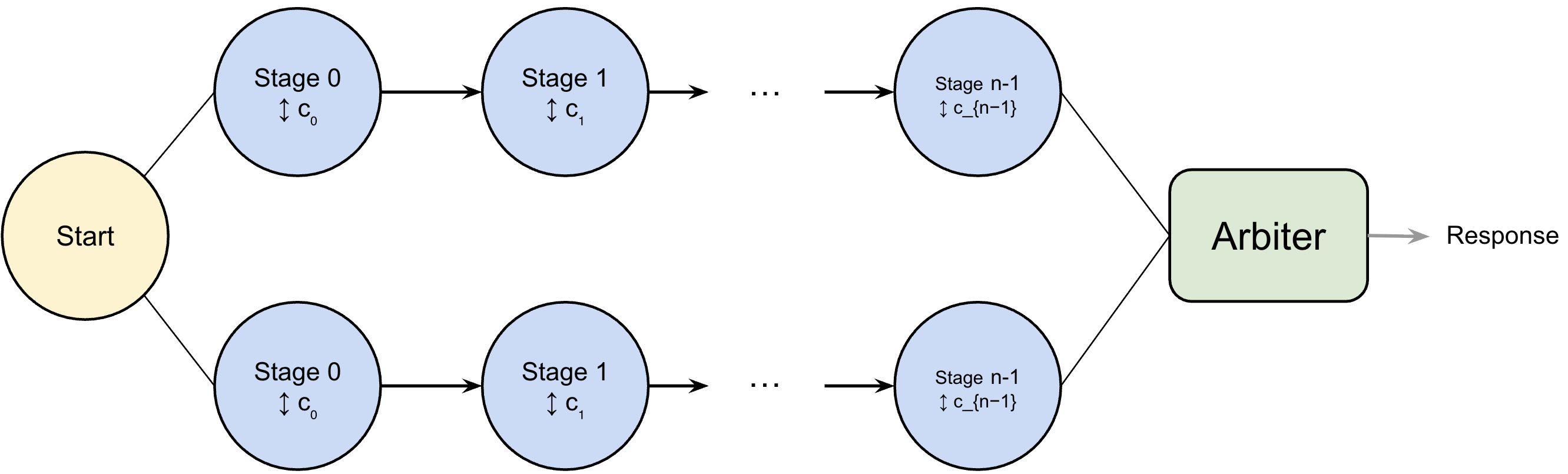}
\caption{Single Arbiter PUF architecture.}
\label{fig:sa_puf}
\vspace{-0.4em}
\end{figure}

As seen in Fig.~\ref{fig:sa_puf} the SA-PUF serves as a baseline architecture. It consists of $N$ cascaded delay stages controlled by challenge bits. A start pulse launches complementary signals into symmetric multiplexed delay paths, where each stage conditionally swaps the signal ordering based on the applied challenge bit. The final race is resolved by an arbiter to produce a response bit \cite{suh2007physical, hemavathy2023arbiter}. Its linear additive delay model makes it vulnerable to modeling attacks \cite{ruhrmair2010modeling, becker2015gap}, providing a reference point for evaluating how increased complexity affects Trojan detectability and stability.


\subsubsection{Feedforward Arbiter PUF (FA-PUF)}
\begin{figure}[H]
\centering
\includegraphics[width=0.9\columnwidth]{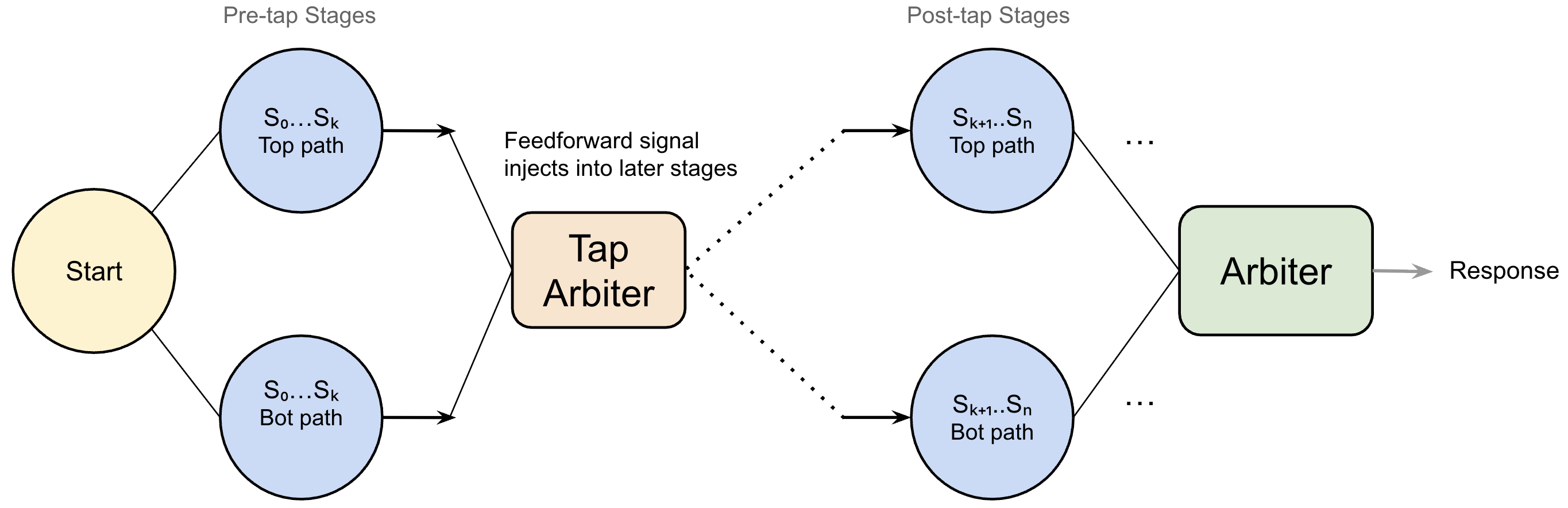}
\caption{Feedforward Arbiter PUF architecture.}
\label{fig:ff_puf}
\vspace{-0.5em}
\end{figure}

The FA-PUF, shown in Fig.~\ref{fig:ff_puf}, introduces stage level dependency by re-injecting an intermediate race result from an early delay stage into later stages through XOR challenge perturbation. This feedforward signal modifies the effective challenge vector applied to subsequent multiplexers, breaking linear separability of the additive delay model. Feedforward structures improve modeling resistance \cite{ebrahimabadi2021novel, hemavathy2023arbiter}, although added complexity does not eliminate vulnerability \cite{tobisch2015scaling}. The FA-PUF serves as an intermediate complexity design for evaluating how structural feedback influences Trojan behavior.

\subsubsection{XOR Arbiter PUF (XA-PUF)}
\begin{figure}[H]
\centering
\includegraphics[width=0.9\columnwidth]{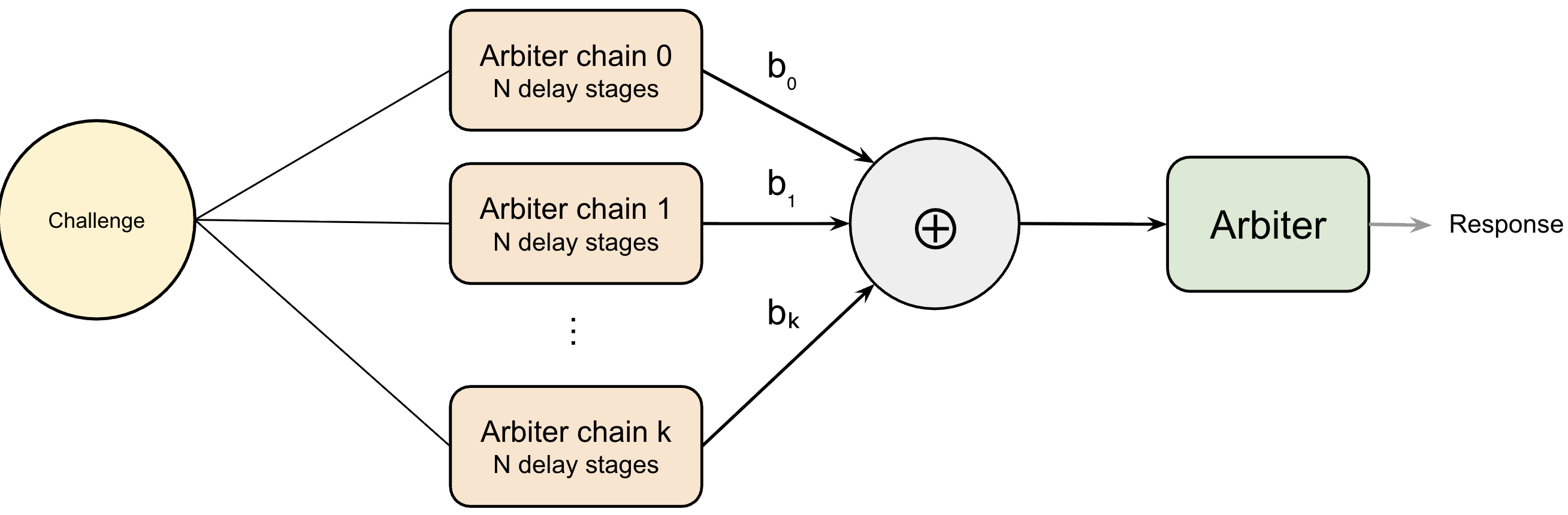}
\caption{XOR Arbiter PUF architecture.}
\label{fig:xor_puf}
\vspace{-0.5em}
\end{figure}

The XA-PUF as seen in Fig.~\ref{fig:xor_puf}, instantiates multiple independent arbiter delay chains in parallel and combines their response bits using bitwise XOR operation, producing a parity-based response. This implementation introduces nonlinear interaction between multiple delay races, and expands the hypothesis space required for modeling \cite{becker2015gap, zhou2017secure}, although large-scale attacks remain effective under realistic conditions \cite{tobisch2015scaling, wisiol2019breaking}. From a Trojan perspective, parallel composition increases activity and logic footprint, allowing evaluation of how aggregation affects detectability.

\subsubsection{Obfuscated Composite Feedforward XOR Interpose Arbiter PUF (OA-PUF)}
\begin{figure}[H]
\centering
\includegraphics[width=0.9\columnwidth]{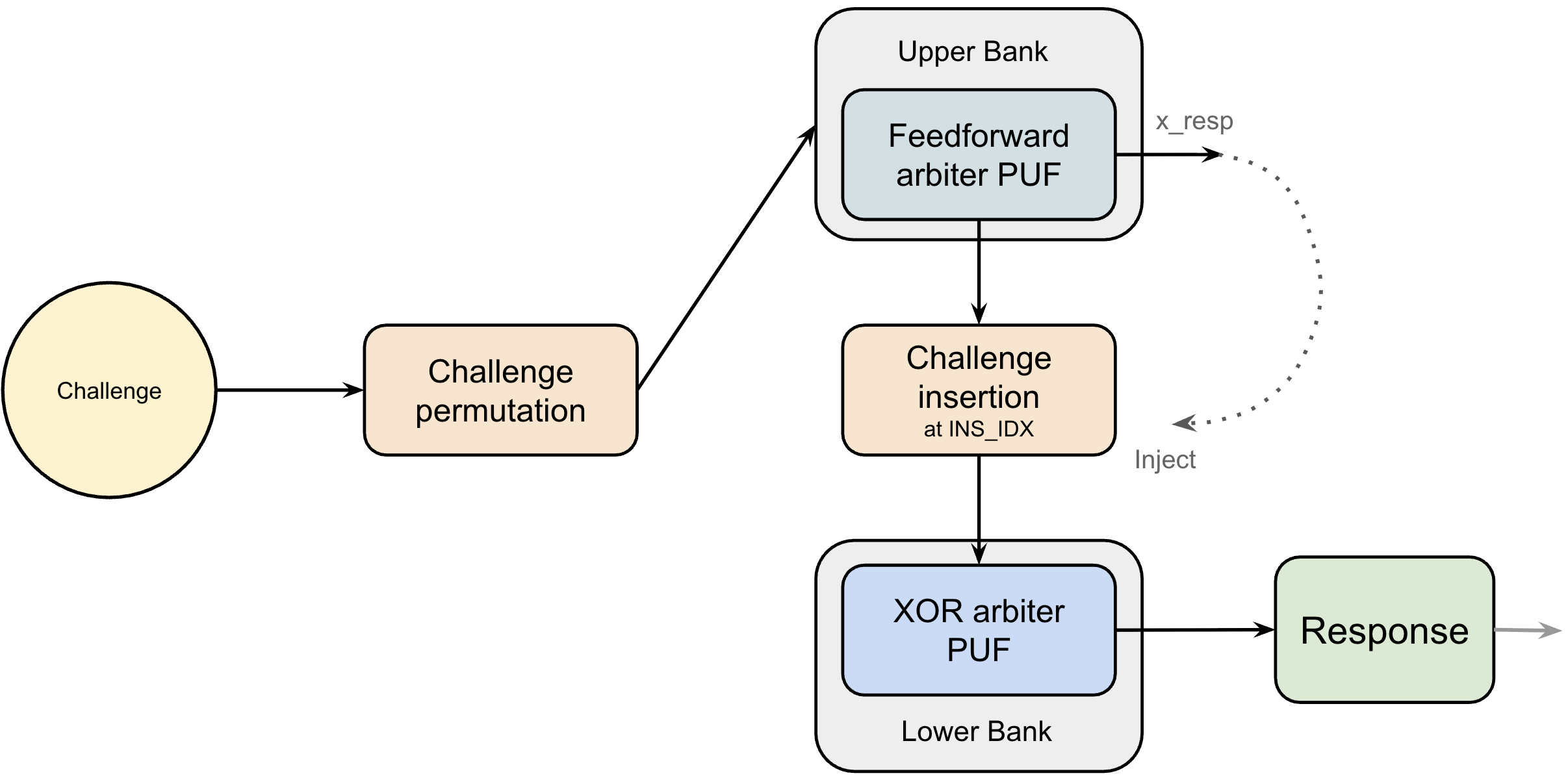}
\caption{Obfuscated Composite Feedforward XOR Interpose Arbiter PUF architecture.}
\label{fig:ocfxi_puf}
\vspace{-0.5em}
\end{figure}

The OA-PUF pictured in Fig.~\ref{fig:ocfxi_puf}, is the most complex architecture in the framework. An upper feedforward PUF bank generates an intermediate response that is interposed into the challenge vector of a lower XOR based bank, forming a hierarchical dependency similar to interpose constructions proposed for modeling resistance \cite{nguyen2018interpose, wisiol2020splitting}. This structure significantly increases internal dependency depth and switching activity, making it useful for evaluating whether architectural complexity conceals malicious logic.

\subsection{Trojans and Triggers}

The Trojans evaluated in this work are manually designed for controlled experimentation. Such designs are based on commonly studied trigger mechanisms described in prior hardware Trojan literature, including sequential, counter, and combinational triggers.

\subsubsection{Sequential}
This Trojan triggers only after a predefined ordered sequence of events. A finite-state machine advances when three specific challenges appear in the correct order, and resets upon deviation \cite{wang2011sequential}. Once the final state is reached, activation is latched and the payload executes. Since activation depends on sequences rather than isolated inputs, accidental triggering is unlikely. This trigger evaluates how rare activation interacts with architectural complexity.

\subsubsection{Counter}
This Trojan activates after prolonged use by counting valid events. In this case, each start pulse increments an internal counter, and activation occurs once a predefined threshold is reached \cite{liu2011design, tehranipoor2010survey}. Because activation depends only on event count, it avoids direct statistical correlation with challenge content. The counter trigger enables evaluation of long latency activation, and whether accumulated usage produces detectable anomalies.

\subsubsection{Hash}
This Trojan activates when an XOR based hash of the input challenge matches a predefined target value. Rather than monitoring raw challenge bits, the trigger transforms challenge segments and compares the result to a hidden constant \cite{xue2020ten}. Unlike sequential triggers, hash-based activation obscures conditions through functional transformation. This nonlinearity enables evaluation of whether increased structural complexity masks observable deviations when concealed conditions are satisfied.

\subsection{Payloads}

\subsubsection{Bias Injection}
This payload forces the PUF output to a fixed logic value after activation, introducing deterministic bias into the challenge–response behavior. Activation occurs on the first valid evaluation following trigger assertion, after which all responses become constant and independent of the delay race result. This degrades uniformity and uniqueness while preserving internal switching activity, potentially limiting detectability through structural or power-based analysis. The payload evaluates whether statistical degradation can remain concealed, particularly in nonlinear architectures.

\subsubsection{Freeze Output}
This payload stores the first valid response after activation and replays it for all subsequent evaluations. Internal delay computation continues, but its result is no longer externally propagated. By eliminating challenge dependent variability, freezing output violates the PUF  requirement of unique CRP mappings \cite{armknecht2016towards}. However, because internal logic continues switching, structural or power based anomalies tend to remain limited. This payload evaluates whether architectural complexity obscures detection of entropy collapse, especially in nonlinear designs where internal activity persists.

\section{Experimental Setup}
This section outlines the experimental methodology for evaluating PUF designs.
\subsection{Simulation Flow}
All experiments are driven by a unified Verilog testbench. Challenge vectors are pre-generated as 64-bit hexadecimal values and stored in external text files. During simulation, these values are loaded into memory and sequentially applied to one or more parameterized PUF wrapper instances.

Each wrapper selected the PUF architecture, Trojan type, payload, and randomness mode at compile time. The testbench generates a one-cycle start pulse per challenge and logged the resulting response along with activation flags and configuration metadata.

Two operating modes are supported: (1) Standard Mode, full logging of responses and configuration fields for comprehensive understanding, and (2) ML Mode, minimal logging of CRPs to generate datasets for model training.

\subsection{Functional Metric Analysis}
Raw simulation logs are exported as CSV files. Metric computation (uniqueness, reliability, uniformity, randomness) is performed using Python scripts that parse the logged data and compute the necessary statistics according to the defined formulas. Reliability experiments utilize two independent simulation runs with identical configuration parameters.

\subsection{Synthesis Analysis}
Each clean and Trojan-infected configuration is synthesized in Vivado. A separate custom synthesis project is used for convenience while maintaining identical implementations, settings, and device parameters. Test cases are selected through a wrapper module, allowing each PUF and Trojan combination to be compiled independently before using the standard Vivado synthesis tool. Resource utilization values are extracted directly from the Vivado synthesis reports, while power estimates are obtained using the Vivado power analysis tool after synthesis. All values are collected from these reports, ensuring that differences across configurations are caused only by the inserted Trojan logic.

\subsection{ML Modeling Attack}
For modeling analysis, 50,000 CRPs are generated per configuration using RTL simulation, specifically in ML mode within the custom testbench. Exported datasets are processed in Python, transformed into the $\Phi$ feature space for modeling, and evaluated using logistic regression. Training and testing splits are performed using fixed random seeds to ensure consistency and reproducibility across experiments. All data is collected from implementations operating at the RTL; therefore, responses reflect purely structural logic behavior rather than silicon delay asymmetries. The modeling analysis focuses on Trojan-induced functional deviations and post-trigger learnability, rather than the physical modeling resistance of the PUFs themselves.

\subsection{Automation and Reproducibility}
All configuration parameters are defined at compile time in the wrapper and testbench. Logging, dataset generation, and file handling are automated to ensure consistency across experimental runs.

\section{Experimental Results}
This section evaluates the functional, structural, and security characteristics of delay‑based PUF architectures under both clean and Trojan‑infected conditions. The source code is available on GitHub \footnote{https://github.com/cars-lab-repo/TroPUF}.

\subsection{Functional Analysis}

\subsubsection{Uniqueness}
Table~\ref{tab:uniqueness} reports uniqueness values across all configurations. Prior to activation, all architectures remain near the ideal 50\% target. Clean and dormant Trojan designs are statistically indistinguishable, confirming that embedded trigger logic does not affect inter-instance variation under normal operation. Architectural complexity does not materially influence baseline uniqueness. After Trojan activation, uniqueness degrades sharply across all architectures. Triggered payload behavior increases response correlation, reducing inter-instance differentiation. This effect is consistent across simple and composite designs, indicating that structural complexity does not prevent post-activation degradation.

    \begin{table}[H]
    \centering
    \caption{Uniqueness Values Across PUF Implementations (\%)}
    \label{tab:uniqueness}
    \scriptsize
    \setlength{\tabcolsep}{3pt}
    \renewcommand{\arraystretch}{0.95}
    \begin{tabular}{|l|c|c|c|c|}
    \hline
    \textbf{Configuration} & \textbf{SA-PUF} & \textbf{FA-PUF} & \textbf{XA-PUF} & \textbf{OA-PUF} \\
    \hline
     \hline    
    Clean v. Clean & 48.9 & 49.6 & 50.2 & 49.8 \\
    \hline
    Clean v. Sequential Trojan (Dormant) & 48.9 & 49.6 & 50.2 & 49.9 \\
    Clean v. Sequential Trojan (Triggered) & 1.2 & 0.9 & 0.6 & 0.4 \\
    \hline
    Clean v. Hash Trojan (Dormant) & 48.9 & 49.6 & 50.2 & 49.9 \\
    Clean v. Hash Trojan (Triggered) & 1.1 & 0.8 & 0.5 & 0.3 \\
    \hline
    Clean v. Counter Trojan (Dormant) & 48.9 & 49.6 & 50.2 & 49.9 \\
    Clean v. Counter Trojan (Triggered) & 1.3 & 1.0 & 0.7 & 0.5 \\
    \hline
    \end{tabular}
    \end{table}

\subsubsection{Reliability}
Table~\ref{tab:reliability} reports reliability results across all configurations. Reliability is evaluated by executing two independent simulation runs using identical challenge sets and configurations, and comparing the responses produced in each run. Results report the reliability between the separate runs. Under clean and dormant conditions, all architectures exhibit perfect reliability, with responses remaining consistent across repeated evaluations. Dormant Trojan logic does not disrupt response stability. Following activation, reliability drops to approximately 50\% across all architectures. Triggered payload behavior interferes with deterministic response generation, producing instability under identical challenges. This degradation occurs uniformly across architectures, indicating that structural strengthening does not mitigate reliability loss once malicious logic is active.

    \begin{table}[H]
    \centering
    \caption{Reliability Values Across PUF Implementations (\%)}
    \label{tab:reliability}
    \scriptsize
    \setlength{\tabcolsep}{3pt}
    \renewcommand{\arraystretch}{0.95}
    \begin{tabular}{|l|c|c|c|c|}
    \hline
    \textbf{Configuration} & \textbf{SA-PUF} & \textbf{FA-PUF} & \textbf{XA-PUF} & \textbf{OA-PUF} \\
    \hline
    \hline
    Clean v. Clean & 100 & 100 & 100 & 100 \\
    \hline
    Clean v. Sequential Trojan (Dormant) & 100 & 100 & 100 & 100 \\
    Clean v. Sequential Trojan (Triggered) & 49.8 & 50.2 & 49.5 & 50.1 \\
    \hline
    Clean v. Hash Trojan (Dormant) & 100 & 100 & 100 & 100 \\
    Clean v. Hash Trojan (Triggered) & 50.3 & 49.7 & 50.0 & 49.6 \\
    \hline
    Clean v. Counter Trojan (Dormant) & 100 & 100 & 100 & 100 \\
    Clean v. Counter Trojan (Triggered) & 49.9 & 50.4 & 49.6 & 50.2 \\
    \hline
    \end{tabular}
    \end{table}

\subsubsection{Uniformity}
Table~\ref{tab:uniformity} reports uniformity values across all configurations. Clean and dormant designs maintain response distributions near the ideal 50\% balance, with no measurable bias introduced prior to activation. Architectural variation does not significantly affect uniformity. After activation, uniformity shifts toward extreme bias across all architectures. Triggered payload behavior constrains output distributions, driving responses away from the ideal state. This effect is consistent regardless of structural complexity.

    \begin{table}[H]
    \centering
    \caption{Uniformity Values Across PUF Implementations (\%)}
    \label{tab:uniformity}
    \scriptsize
    \setlength{\tabcolsep}{3pt}
    \renewcommand{\arraystretch}{0.95}
    \begin{tabular}{|l|c|c|c|c|}
    \hline
     \textbf{Configuration} & \textbf{SA-PUF} & \textbf{FA-PUF} & \textbf{XA-PUF} & \textbf{OA-PUF} \\
    \hline
    \hline
    Clean & 48.7 & 49.4 & 50.1 & 49.8 \\
    \hline
    Sequential Trojan (Dormant) & 48.7 & 49.4 & 50.1 & 49.8 \\
    Sequential Trojan (Triggered) & 99.2 & 99.5 & 99.7 & 99.9 \\
    \hline
    Hash Trojan (Dormant) & 48.7 & 49.4 & 50.1 & 49.8 \\
    Hash Trojan (Triggered) & 99.1 & 99.6 & 99.8 & 99.9 \\
    \hline
    Counter Trojan (Dormant) & 48.7 & 49.4 & 50.1 & 49.8 \\
    Counter Trojan (Triggered) & 99.3 & 99.4 & 99.7 & 99.8 \\
    \hline
    \end{tabular}
    \end{table}

\subsubsection{Randomness}
  
    Table~\ref{tab:randomness} reports randomness (entropy) values across all configurations. Under clean and dormant conditions, all architectures exhibit near-maximum entropy, indicating statistically unpredictable behavior. No difference is observed between clean and Trojan-infected designs prior to activation. Following activation, entropy drops sharply across all architectures. Triggered payload behavior reduces response variability and increases predictability. This degradation occurs uniformly across designs, demonstrating that architectural obfuscation does not preserve unpredictability once malicious logic becomes active.

    \begin{table}[H]
    \centering
    \caption{Randomness Values Across PUF Implementations (\%)}
    \label{tab:randomness}
    \scriptsize
    \setlength{\tabcolsep}{3pt}
    \renewcommand{\arraystretch}{0.95}
    \begin{tabular}{|l|c|c|c|c|}
     \hline
      \textbf{Configuration} & \textbf{SA-PUF} & \textbf{FA-PUF} & \textbf{XA-PUF} & \textbf{OA-PUF} \\
    \hline
    \hline
    Clean & 99.89 & 99.99 & 99.98 & 99.99 \\
    \hline
    Sequential Trojan (Dormant) & 99.89 & 99.99 & 99.98 & 99.99 \\
    Sequential Trojan (Triggered) & 2.56 & 1.98 & 0.87 & 0.15 \\
    \hline
    Hash Trojan (Dormant) & 99.99 & 99.99 & 99.98 & 99.99 \\
    Hash Trojan (Triggered) & 1.47 & 0.92 & 0.41 & 0.12 \\
    \hline
    Counter Trojan (Dormant) & 99.99 & 99.99 & 99.98 & 99.99 \\
    Counter Trojan (Triggered) & 1.43 & 1.01 & 0.76 & 0.20 \\
    \hline
    \end{tabular}
    \end{table}

\subsection{Synthesis Analysis}
\subsubsection{Lookup Table Utilization}
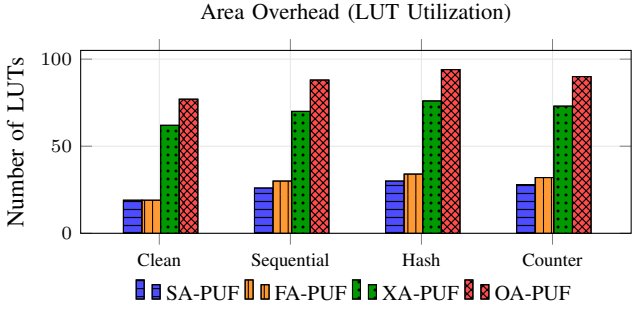
\begin{figure}[H]
\centering
\begin{tikzpicture}
\begin{axis}[
    ybar,
    width=\linewidth,
    height=4cm,
    bar width=7pt,
    title={Area Overhead (LUT Utilization)},
    title style={font=\small, yshift=1pt},
    ylabel={Number of LUTs},
    ymin=0,
    ymax=105,
    xtick=data,
    symbolic x coords={
        Clean,
        SeqDorm,
        HashDorm,
        CtrDorm
    },
    xticklabels={
        Clean,
        {Sequential},
        {Hash},
        {Counter}
    },
    tick label style={font=\scriptsize},
    xticklabel style={align=center},
    enlarge x limits=0.2,
    legend style={
        at={(0.5,-0.22)},
        anchor=north,
        legend columns=4,
        draw=none,
        font=\footnotesize
    },
    grid=major,
    grid style={gray!20}
]
\addplot[
fill=blue!70,
draw=black,
postaction={pattern=horizontal lines, pattern color=black},
bar shift=-10pt
] coordinates {
    (Clean,19)
    (SeqDorm,26)
    (HashDorm,30)
    (CtrDorm,28)
};
\addplot[
fill=orange!80,
draw=black,
postaction={pattern=vertical lines, pattern color=black},
bar shift=-3pt
] coordinates {
    (Clean,19)
    (SeqDorm,30)
    (HashDorm,34)
    (CtrDorm,32)
};
\addplot[
fill=green!60!black,
draw=black,
postaction={pattern=dots, pattern color=black},
bar shift=4pt
] coordinates {
    (Clean,62)
    (SeqDorm,70)
    (HashDorm,76)
    (CtrDorm,73)
};
\addplot[
fill=red!70,
draw=black,
postaction={pattern=crosshatch, pattern color=black},
bar shift=11pt
] coordinates {
    (Clean,77)
    (SeqDorm,88)
    (HashDorm,94)
    (CtrDorm,90)
};

\legend{SA-PUF, FA-PUF, XA-PUF, OA-PUF}

\end{axis}
\end{tikzpicture}
\caption{Lookup Table (LUT) utilization across PUF architectures.}
\label{fig:lut_utilization}
\end{figure}

Fig.~\ref{fig:lut_utilization} reports LUT utilization across all configurations. Trojan insertion introduces additional combinational logic due to trigger and control circuitry. However, the increases remain slight and consistent across Trojan types. Relative overhead varies by architecture. Simpler designs exhibit greater sensitivity, while more complex architectures absorb the additional logic with reduced relative impact. Overall, LUT-based analysis alone provides limited discriminatory power for detecting dormant Trojans, particularly in designs with higher inherent complexity.

\subsubsection{Flip-Flop Utilization}
\begin{figure}[H]
\centering
\begin{tikzpicture}
\begin{axis}[
    ybar,
    width=\linewidth,
    height=4cm,
    bar width=7pt,
    title={Area Overhead (Flip-Flop Utilization)},
    title style={font=\small, yshift=1pt},
    ylabel={Number of Flip-Flops},
    ymin=0,
    ymax=16,
    xtick=data,
    symbolic x coords={
        Clean,
        SeqDorm,
        HashDorm,
        CtrDorm
    },
    xticklabels={
        Clean,
        {Sequential},
        {Hash},
        {Counter}
    },
    tick label style={font=\scriptsize},
    xticklabel style={align=center},
    enlarge x limits=0.2,
    legend style={
        at={(0.5,-0.22)},
        anchor=north,
        legend columns=4,
        draw=none,
        font=\footnotesize
    },
    grid=major,
    grid style={gray!20}
]
\addplot[
fill=blue!70,
draw=black,
postaction={pattern=horizontal lines, pattern color=black},
bar shift=-10pt
] coordinates {
    (Clean,1)
    (SeqDorm,4)
    (HashDorm,3)
    (CtrDorm,6)
};

\addplot[
fill=orange!80,
draw=black,
postaction={pattern=vertical lines, pattern color=black},
bar shift=-3pt
] coordinates {
    (Clean,2)
    (SeqDorm,5)
    (HashDorm,4)
    (CtrDorm,7)
};

\addplot[
fill=green!60!black,
draw=black,
postaction={pattern=dots, pattern color=black},
bar shift=4pt
] coordinates {
    (Clean,5)
    (SeqDorm,8)
    (HashDorm,7)
    (CtrDorm,10)
};

\addplot[
fill=red!70,
draw=black,
postaction={pattern=crosshatch, pattern color=black},
bar shift=11pt
] coordinates {
    (Clean,9)
    (SeqDorm,12)
    (HashDorm,11)
    (CtrDorm,14)
};

\legend{SA-PUF, FA-PUF, XA-PUF, OA-PUF}

\end{axis}
\end{tikzpicture}
\caption{Flip-Flop (FF) utilization across PUF architectures.}
\label{fig:ff_utilization}
\end{figure}
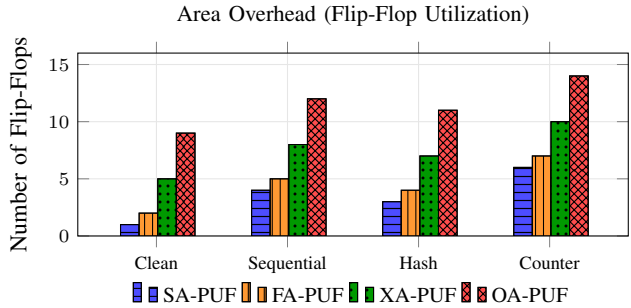

Fig.~\ref{fig:ff_utilization} reports flip-flop utilization across all configurations. Register usage increases primarily, however slightly, in stateful trigger implementations, such as counter mechanisms, while stateless triggers introduce smaller overhead. The relative impact depends on baseline sequential complexity. Architectures with minimal native register usage exhibit more visible proportional increases, whereas more complex designs absorb the additional registers with reduced relative effect. Consequently, register based analysis provides limited effectiveness for identifying dormant Trojans.

\subsubsection{Power Consumption}
\begin{figure}[!ht]
\centering
\begin{tikzpicture}
\begin{axis}[
    ybar,
    width=\linewidth,
    height=4cm,
    bar width=7pt,
    title={Total On-Chip Power (W)},
    title style={font=\small, yshift=1pt},
    ylabel={Power (W)},
    ymin=1.8,
    ymax=3.2,
    xtick=data,
    symbolic x coords={
        Clean,
        SeqDorm,
        HashDorm,
        CtrDorm
    },
    xticklabels={
        Clean,
        {Sequential\\(Dormant)},
        {Hash\\(Dormant)},
        {Counter\\(Dormant)}
    },
    tick label style={font=\scriptsize},
    xticklabel style={align=center},
    enlarge x limits=0.2,
    legend style={
        at={(0.5,-0.35)},
        anchor=north,
        legend columns=4,
        draw=none,
        font=\footnotesize
    },
    grid=major,
    grid style={gray!20}
]
\addplot[
fill=blue!70,
draw=black,
postaction={pattern=horizontal lines, pattern color=black},
bar shift=-10pt
] coordinates {
    (Clean,1.95)
    (SeqDorm,2.05)
    (HashDorm,2.10)
    (CtrDorm,2.02)
};

\addplot[
fill=orange!80,
draw=black,
postaction={pattern=vertical lines, pattern color=black},
bar shift=-3pt
] coordinates {
    (Clean,2.05)
    (SeqDorm,2.10)
    (HashDorm,2.18)
    (CtrDorm,2.12)
};

\addplot[
fill=green!60!black,
draw=black,
postaction={pattern=dots, pattern color=black},
bar shift=4pt
] coordinates {
    (Clean,2.78)
    (SeqDorm,2.70)
    (HashDorm,2.76)
    (CtrDorm,2.72)
};

\addplot[
fill=red!70,
draw=black,
postaction={pattern=crosshatch, pattern color=black},
bar shift=11pt
] coordinates {
    (Clean,2.87)
    (SeqDorm,2.92)
    (HashDorm,3.02)
    (CtrDorm,2.90)
};

\legend{SA-PUF, FA-PUF, XA-PUF, OA-PUF}

\end{axis}
\end{tikzpicture}
\caption{Total on-chip power across PUF architectures.}
\label{fig:power_utilization}
\end{figure}
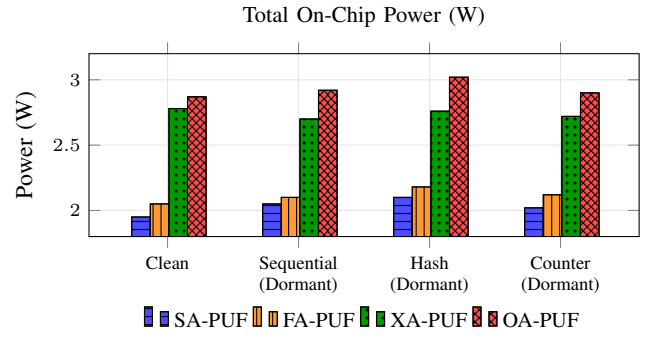

Fig.~\ref{fig:power_utilization} reports total on-chip power estimates across configurations. Under dormant conditions, power values remain within expected estimation noise, with no consistent or monotonic relationship observed across Trojan types. Dormant Trojan logic does not produce a distinguishable power signature. Any incremental consumption is masked by the surrounding switching activity of the PUF circuitry. Architectures with higher inherent activity further reduce observable impact, limiting the effectiveness of power-based detection.

\subsection{ML Modeling Attack}
Table~\ref{tab:ml_accuracy} reports logistic regression prediction accuracy across implementations of all evaluated PUF architectures. All clean and dormant configurations showcase prediction accuracy around 50\%, corresponding to near random guessing. This behavior is expected considering experiments are conducted at the RTL level. As a result, the PUF responses are statistically balanced and contain no structural bias that a learning algorithm can exploit.

Dormant Trojan insertion does not alter modeling accuracy across any architecture. Clean and infected (pre-trigger) implementations remain indistinguishable. This indicates that inactive Trojan logic does not introduce observable artifacts at the structural level that can be learned by a model.

In contrast, once Trojans are triggered, prediction accuracy increases dramatically to above 99\%. Trojan activation exploits entropic PUF behavior, making responses predictable. This transition confirms that Trojan presence fundamentally alters the functional characteristics of the PUFs only after activation, while remaining completely stealthy prior to triggering, proving that while dormant, they can not be detected via modeling.

\begin{table}[H]
    \centering
    \caption{ML Model Prediction Accuracy Across PUF Implementations (\%)}
    \label{tab:ml_accuracy}
    \scriptsize
    \setlength{\tabcolsep}{3pt}
    \renewcommand{\arraystretch}{0.95}
    \begin{tabular}{|l|c|c|c|c|}
    \hline
     \textbf{Configuration} & \textbf{SA-PUF} & \textbf{FA-PUF} & \textbf{XA-PUF} & \textbf{OA-PUF} \\
    \hline
    \hline
    Clean & 49.3 & 51.3 & 49.6 & 49.7 \\
    \hline
    Sequential Trojan (Dormant) & 49.3 & 51.3 & 49.6 & 49.7 \\
    \hline
    Hash Trojan (Dormant) & 49.3 & 51.3 & 49.6 & 49.7 \\
    \hline
    Counter Trojan (Dormant) & 49.3 & 51.3 & 49.6 & 49.7 \\
    \hline
    Sequential Trojan (Triggered) & 99.5 & 99.7 & 99.8 & 99.7 \\
    \hline
    Hash Trojan (Triggered) & 99.7 & 99.7 & 99.8 & 99.9 \\
    \hline
    Counter Trojan (Triggered) & 99.4 & 99.6 & 99.7 & 99.6 \\
    \hline
    \end{tabular}
\end{table}

\section{Conclusions}
In this work, we showed that hardware Trojans could be successfully embedded within PUF architectures while remaining undetectable. Dormant Trojan‑infected designs exhibited no meaningful behavioral differences compared to clean instances, but once activated, Trojans caused immediate changes in PUF behavior. 

This reveals a fundamental limitation in current PUF safety and validation methodologies. In short, satisfying the regular criteria does not imply trustworthiness. The same characteristics that make PUFs effective security primitives can also enable them to conceal malicious logic, thereby undermining the trust they are intended to provide.  A PUF can appear statistically ideal, yet still contain malicious insertions. As a result, existing Trojan detection approaches and PUF validation pipelines are insufficient when applied in isolation. New detection strategies must move beyond current assumptions, and address the presence of malicious logic embedded directly within the security primitive itself. 

\balance
\bibliographystyle{plain}
\bibliography{references}

\end{document}